\documentclass[a4paper,11pt,english]{amsart}
\def\margin_comment#1{\marginpar{\sffamily{\tiny #1\par}\normalfont}}
\usepackage[table]{xcolor}
\usepackage{tikz}
\usepackage[braket, qm]{qcircuit}

\usepackage{amssymb,euscript}
\usepackage{amsfonts}
\usepackage{arydshln}
\usepackage{dutchcal}
\usepackage{marginnote}

\usepackage{graphicx}
\usepackage{setspace}
\usepackage{color}
\usepackage{amsmath,amssymb,amsthm}
\usepackage{epsfig}
\usepackage{babel}
\usepackage{tikz}
 \usepackage{amsmath}
\usepackage{tikz-cd}
\usetikzlibrary{matrix, calc, arrows}
\usepackage{tikz}
\usetikzlibrary{arrows,chains,matrix,positioning,scopes}
\usepackage{lipsum}
\usepackage{mathtools}
\usepackage{mathrsfs}
\usepackage{graphicx}
\usepackage{graphics}
\usepackage{setspace}
	\usepackage{color}

\usetikzlibrary{trees}
\usetikzlibrary{arrows,shapes,automata,backgrounds,petri,positioning,scopes,matrix, calc, plothandlers,plotmarks,patterns}
\usepackage{enumerate}
\usepackage{float}
\usepackage{tikz-3dplot}
\usepackage{pgfplots}
\usepackage{tkz-graph}
\usepackage[top=1.5cm, bottom=1.5cm, outer=5cm, inner=2cm, heightrounded, marginparwidth=2.5cm, marginparsep=2cm]{geometry}

\makeatother

\newcommand{\scalemath}[2]{\scalebox{#1}{\begin{math} {#2} \end{math}}}

\newcommand*{\mat}{\mathbf}

\newcommand{\rvline}{\hspace*{-\arraycolsep}\vline\hspace*{-\arraycolsep}}
\newgeometry{left=3cm,right=3cm,top=3.5cm,bottom=3cm}

\GraphInit[vstyle = Normal]
\tikzset{
  LabelStyle/.style = {minimum width = 2em, 
                        text = red, font = \bfseries },
  VertexStyle/.append style = { inner sep=2pt,
                                font = \Large\bfseries, fill},
  EdgeStyle/.append style = {->, bend left} }

\makeatletter
\tikzset{join/.code=\tikzset{after node path={%
\ifx\tikzchainprevious\pgfutil@empty\else(\tikzchainprevious)%
edge[every join]#1(\tikzchaincurrent)\fi}}}
\makeatother
\tikzset{>=stealth',every on chain/.append style={join},
         every join/.style={->}}
\tikzstyle{labeled}=[execute at begin node=$\scriptstyle,
   execute at end node=$]
\newtheorem{thm}{Theorem}[section]
\numberwithin{equation}{section} 
\numberwithin{figure}{section} 
\theoremstyle{plain}
\newtheorem*{thm*}{Theorem}
\theoremstyle{definition}
\theoremstyle{plain}
\newtheorem{thmx}{Theorem}
 
\newtheorem*{defn*}{Definition}

\theoremstyle{plain}

\theoremstyle{plain} 

\theoremstyle{plain}

\newtheorem{prop}[thm]{Proposition} 
\theoremstyle{remark}
\newtheorem{ex}[thm]{Example}
\theoremstyle{remark}

\theoremstyle{plain}

\theoremstyle{plain}

\theoremstyle{plain}
\newtheorem{lem}[thm]{Lemma} 
\theoremstyle{definition}
\newtheorem{defn}[thm]{Definition}

\newtheorem*{acknowledgment*}{Addentum}

\theoremstyle{plain}
\newtheorem*{ex*}{Example}
\theoremstyle{plain}

\begin{document}
\title{On the Realization of quantum gates coming from the Tracy-Singh product}
\author{Fabienne Chouraqui}

\begin{abstract}
 The Tracy-Singh product of matrices permits to construct a new gate   $c \boxtimes c'$ from two $2$-qudit gates $c$ and $c'$. If  $c$ and $c'$ are both Yang-Baxter gates, then  $c \boxtimes c'$ is also a Yang-Baxter gate, and if at least one of them is entangling, then $c \boxtimes c'$ is also entangling. A  natural question arises about  the realisation of  these gates, $c \boxtimes c'$, in terms of  local and universal gates. In this paper, we consider this question and describe this realisation.
\end{abstract}
\maketitle
\section*{Introduction}
 Let $V$ be   a finite-dimensional vector space over the  field $\mathbb{C}$.  The linear operator  $c:V \otimes V \rightarrow V \otimes V$    is a solution of the  Yang-Baxter equation (YBE) if it  satisfies  the equality  $c^{12}c^{23}c^{12}=c^{23}c^{12}c^{23}$ in $V \otimes V \otimes V$, $c$ is also called a braiding operator or an $R$-matrix.  
 Every  unitary solution of  the YBE, $c:V \otimes V \rightarrow V \otimes V$, where $V$ is a complex vector space of dimension $d$, can be seen as a  2-qudit gate $c:\,(\mathbb{C}^d)^{\otimes 2}\,\rightarrow\,(\mathbb{C}^d)^{\otimes 2}$.  
 
  The Yang-Baxter equation plays a significant role in quantum computing, particularly in the context of quantum information theory and in the domain of  integrable quantum computation,  a type of quantum circuit model of computation in which two-qubit gates are either the swap gate or non-trivial unitary solutions of the YBE \cite{integrable-quantum}.
 Recently, the YBE has been used in the study of quantum error correction   \cite{ybe-quantum-correction}, and in teleportation-based quantum computation \cite{teleportation}.  There is also an increasing interest on the realisation and the implementation of  Yang–Baxter gates and the Yang–Baxter equation on quantum computers. The Yang–Baxter gates from the integrable circuit have been implemented on IBM superconducting-based quantum computers \cite{imp1,imp2}, and  the YBE has been tested on the optical and NMR systems \cite{test1,test2}. In \cite{implement}, the authors address the question of how to find  optimal  realisations of Yang-Baxter gates. 
 
  In \cite{kauf-lo3}, L.H. Kaufman and S.J. Lomonaco Jr  initiate the study of entangling $2$-qubit gates, $R:\,\mathbb{C}^2\otimes\mathbb{C}^2\,\rightarrow\,\mathbb{C}^2\otimes\mathbb{C}^2$, which  satisfy  the YBE.  In \cite{chou-quantum}, it is shown that for every integer $d\geq 2$,  there exists an entangling  $2$-qudit gate  $R:\,(\mathbb{C}^d)^{\otimes 2}\,\rightarrow\,(\mathbb{C}^d)^{\otimes 2}$  and there exists a primitive  $2$-qudit gate  $S:\,(\mathbb{C}^d)^{\otimes 2}\,\rightarrow\,(\mathbb{C}^d)^{\otimes 2}$, where  $R$ and $S$ satisfy  the YBE. The proof is by construction and it relies on the application of the Tracy-Singh product of matrices. Indeed,  given $c: (\mathbb{C}^n)^{\otimes 2} \rightarrow (\mathbb{C}^n)^{\otimes 2} $ and $c':(\mathbb{C}^m)^{\otimes 2}\rightarrow (\mathbb{C}^m)^{\otimes 2}$ two   (unitary) solutions of 	the  YBE, with matrices $c$ and $c'$ with respect to their standard  bases respectively, then  the linear operator  $\tilde{c}:(\mathbb{C}^{nm})^{\otimes 2} \rightarrow(\mathbb{C}^{nm})^{\otimes 2}$ with representing matrix  their  Tracy-Singh product with a  canonical block partition, 	$c\boxtimes c'$,   is a (unitary) solution of the YBE \cite{chou-quantum}. In the case that $m=n$, one can consider $\tilde{c}$  as a 2-qudit gate of the form 
  $(\mathbb{C}^{n^2})^{\otimes 2}\,\rightarrow\,(\mathbb{C}^{n^2})^{\otimes 2}$,  or as a 4-qudit gate of the form 
 $(\mathbb{C}^n)^{\otimes 4}\,\rightarrow\,(\mathbb{C}^n)^{\otimes 4}$, and this process can be reiterated to obtain any $2^k$-qudit gate of the form 
 $(\mathbb{C}^n)^{\otimes 2^k}\,\rightarrow\,(\mathbb{C}^n)^{\otimes 2^k}$.

 A collection  $\mathcal{U}$ of $1$-qudit gates $\{S_i\}$ and $2$-qudit gates $\{U_j\}$ is called \emph{exactly universal} if, for each $n \geq  2$, every $n$-qudit gate can be obtained exactly by a circuit made up of the $n$-qudit gates produced by the $\{S_i\}$ and $\{U_j\}$. A set  $\mathcal{U}$ of $1$-qudit gates $\{S_i\}$ and a single  $2$-qudit gate $U$  is exactly universal  if and only if $U$ is an entangling $2$-qudit gate  \cite{bryl}. A question that arises naturally is whether there exists a simple way to realise  a $4$-qudit gate   of the form  $c \boxtimes c'$, or a $2^k$-qudit gate obtained iteratively, in terms of  gates of an  exactly  universal set $\mathcal{U}$. It holds that whenever the realisation  of  the gates $c$ and $c'$  in terms of  gates of   $\mathcal{U}$ is given, the answer is positive.

\begin{thmx}\label{theorem-paper}
Let 	$c, c' :\, (\mathbb{C}^d)^{\otimes 2} \rightarrow (\mathbb{C}^d)^{\otimes 2}$  be $2$-qudit gates. 	Let $c\boxtimes c'$ be the $4$-qudit gate obtained by their  Tracy-Singh product with the  canonical block partition. Let $\mathcal{U}$ be an  exactly universal set of  gates.  If the realisation of  $c$ and $c'$  in terms of the gates in $\mathcal{U}$ is known, then 	the realisation of $c\boxtimes d$ in terms of the gates in $\mathcal{U}$ can be obtained also.  Furthermore, this process can be reiterated to obtain the realisation of any $2^k$-qudit gate of the form 
$(\mathbb{C}^d)^{\otimes 2^k}\,\rightarrow\,(\mathbb{C}^d)^{\otimes 2^k}$.
\end{thmx}
The question whether   a $2$-qudit gate $c$  that satisfy the YBE is  primitive     has found an interesting application in the domain of  link invariants.  Indeed, any solution $c$ of the YBE  yields a family of representations of the braid groups $\rho_n^{c}:\,B_n\rightarrow \operatorname{GL}(V^{\otimes n})$, and V. Turaev  shows in \cite{turaev},   that if $(c,\mu)$ is an enhanced YBE pair, where $\mu \in  \operatorname{End}(V)$, then  the appropriately normalized trace of these representations,  $I_{c}(b)\,=\,\operatorname{Trace}(\rho_n^{c}(b)\cdot\mu^{\otimes n})$, $b \in B_n$,   yields a link invariant. In  \cite{entangle-knot}, the authors prove that  if  such   a  $c$  is  primitive (non-entangling), then the  link invariant obtained from $c$,  using the Turaev construction  \cite{turaev}, is trivial. We show that the  Tracy-Singh product with the  canonical block partition of enhanced  YBE pairs is also an enhanced YBE pair, and  that if $c$ and $c'$ are both primitive of some  kind, then $I_{c\boxtimes c'}$ is  equal to the product of $I_{c}(b)$ and $I_{c'}(b)$. In \cite{chou-quantum}, it is proved that if $c$ and $c'$ are both primitive of the same kind, then $c\boxtimes c'$
is also  primitive (of the same kind), and from \cite{entangle-knot},  $I_{c\boxtimes c'}$ is trivial. Theorem \ref{theorem-paper2} permits to partially recover this result.
\begin{thmx}\label{theorem-paper2}
	Let 	$c, c' :\, (\mathbb{C}^d)^{\otimes 2} \rightarrow (\mathbb{C}^d)^{\otimes 2}$  be $2$-qudit gates, that satisfy the YBE. 	Let $\mu,\eta \,\in \operatorname{End}(\mathbb{C}^d)$, such that $(c,\mu)$ and $(c',\eta)$ are enhanced YBE pairs. Let $c\boxtimes c'$ be the $4$-qudit gate obtained by their  Tracy-Singh product with the  canonical block partition. Then $(c\boxtimes c',\, \mu \otimes \eta)$ is also enhanced YBE. Furthermore, if $c$ and $c'$ are primitive with  $c\,=\,c_1\otimes c_2$ and  $c'\,=\,d_1\otimes d_2$,   $c_1,c_2,d_1,d_2 \in  \operatorname{End}(\mathbb{C}^d)$, then $I_{c\boxtimes c'}(b)\,=\,I_{c}(b)\,\cdot\, I_{c'}(b)$, for every $b \in B_n$. 
\end{thmx}
 The paper is organized as follows. In Section $1$, we give some preliminaries on the YBE, on the Tracy-Singh product of matrices and on  YBE gates constructed  from the Tracy-Singh product. In Section $2$, we discuss the realisation of YBE gates and in particular of  those  constructed  from the Tracy-Singh product.  In Section $3$, we  prove Theorem \ref{theorem-paper2}.  At the end, there is an appendix where  we prove two equations needed in the proof of Theorem \ref{theorem-paper2}.

\section{Yang-Baxter equation and Yang-Baxter gates} 
\subsection{The Yang-Baxter equation (YBE)} 
\begin{defn}\cite[Ch.VIII]{kassel}
	Let $V$ be a vector space over $\mathbb{C}$. A linear automorphism $c$ of  $V \otimes V$ is said to be an $R$-matrix if it is a solution of the Yang-Baxter equation 
	\begin{equation}\label{eqn-ybe}
	(c \otimes Id_V)(Id_V \otimes c )(c \otimes Id_V)\,=\,(Id_V \otimes c)(c \otimes Id_V)(Id_V\otimes c)
	\end{equation}
	that holds in the automorphism group of  $V \otimes V \otimes V$. It is written as $c^{12}c^{23}c^{12}=c^{23}c^{12}c^{23}$. 
\end{defn}
\begin{ex}\label{ex1-kaufman}(\cite{kauf-lo3}, \cite{kassel})
	Let $c,\,d:V \otimes V \rightarrow V \otimes V$ be  $R$-matrices, with $\operatorname{dim}(V)=2$:
	\[c=	\begin{pmatrix}
	\frac{1}{\sqrt{2}}& 0 & 0 & 	\frac{1}{\sqrt{2}}\\
	0 &	\frac{1}{\sqrt{2}} & -	\frac{1}{\sqrt{2}} & 0 \\
	0 &	\frac{1}{\sqrt{2}} & 	\frac{1}{\sqrt{2}} &0 \\
	-	\frac{1}{\sqrt{2}} &0 & 0 &	\frac{1}{\sqrt{2}} \\
	\end{pmatrix}\;\;\;  \textrm{and}\;\;\; 
	d=\begin{pmatrix}
	2& 0 & 0 & 	0\\
	0 &0 & 1 & 0 \\
	0 &	1 & 1.5&0 \\
	0&0 & 0 &	2\\
	\end{pmatrix}\]
	Here $c(e_1 \otimes e_2)=\,\frac{1}{\sqrt{2}} \,e_1 \otimes e_2\,+\,\frac{1}{\sqrt{2}} \,e_2 \otimes e_1$,  $d(e_1 \otimes e_2)=\,e_2 \otimes e_1$, with only $c$  unitary.
\end{ex}
\subsection{Yang-Baxter gates from the Tracy-Singh product} 
The Kronecker product  (or tensor product) of matrices is a fundamental concept in linear algebra and   the Tracy-Singh product of matrices is a  generalisation of it,   called sometimes the block Kronecker product, as  they  share many properties and it requires a block partition of the matrices. Let $A=(a_{ij})$ be a matrix  of size $m \times n$ and $B=(b_{kl})$ of size $p \times q$. Let $A=(A_{ij})$ be partitioned with $A_{ij}$ of size $m_i \times n_j$ as the $ij$-th block submatrix and let $B=(B_{kl})$ be partitioned with $B_{kl}$ of size $p_k \times q_l$ as the $kl$-th block submatrix ($\sum m_i=m,\,\sum n_j=n,\,\sum p_k=p,\,\sum q_l=q$). The  Tracy-Singh (or block Kronecker)  product is defined as follows:
\begin{equation*}
A\,\boxtimes \, B\,=\,((A_{ij} \otimes B_{kl})_{kl})_{ij}
\end{equation*}
The matrix $A\,\boxtimes \, B$ is of size $mp \times nq$ and the block $A_{ij} \otimes B_{kl}$ is size $m_ip \times n_jq$.
For non-partitioned matrices,  $A\,\boxtimes \, B\,=\, A\,\otimes \, B$. We refer to \cite{hyland,neud, liu, tracy-jina,tracy}  for more details and also to \cite{chou-quantum} for details related to the topic of this paper and an example. In the following Theorem, we list  important properties of the Tracy-Singh product.
\begin{thm}\cite{tracy}\label{thm-tracy}
	Let $A$, $B$, $C$, and $D$  be matrices. Then
	\begin{enumerate}[(i)]
		\item $A \boxtimes B$ and $B \boxtimes A$ exist for any  matrices $A$ and $B$.
		\item $A \boxtimes B \neq B \boxtimes A$ in general.
		\item  $(A \boxtimes B)\,\boxtimes C= \,A\boxtimes \,(B \boxtimes C)$.
		\item $(A+ B)\, \boxtimes\,(C +D)=\, A \boxtimes C+A \boxtimes D +B \boxtimes C+B \boxtimes D$,  if $A+B$ and $C+D$ exist.
		\item $(A \boxtimes B)\,(C \boxtimes D)=\, AC \boxtimes BD$,  if $AC$ and $BD$ exist.
		\item $(cA) \boxtimes B\,=\,c (A\boxtimes B \,=\,A \boxtimes (cB)$.
		\item  $(A \boxtimes B)^{-1}\,=  A^{-1} \boxtimes B^{-1}$, if $A$ and $B$ are invertible.
		\item  $(A \boxtimes B)^{t}\,=  A^{t} \boxtimes B^{t}$.
		\item 	$\mat{I}_{n} \boxtimes \mat{I}_{m}\,=\,\mat{I}_{nm}$ for identity partitioned matrices.
	\end{enumerate}
\end{thm}

For any matrix $c$ of size $n^2\times p^2$, there exists  a unique block partition such that all the blocks are matrices of  the same  size, called  \emph{the canonical block partition of $c$}. For the $R$-matrix $c$ from Example \ref{ex1-kaufman}, its  canonical partition into blocks is:
 $\scalemath{0.92}{c=	\begin{pmatrix}
	\frac{1}{\sqrt{2}}& 0 &	 \rvline & 0 & 	\frac{1}{\sqrt{2}}\\
	0 &	\frac{1}{\sqrt{2}} 	& \rvline & -	\frac{1}{\sqrt{2}} & 0 \\
	\hline
	0 &	\frac{1}{\sqrt{2}} 	& \rvline &	\frac{1}{\sqrt{2}} &0 \\
	-	\frac{1}{\sqrt{2}} &0 	& \rvline & 0 &	\frac{1}{\sqrt{2}} \\
	\end{pmatrix}}$.  
\begin{prop}\label{prop-box-cd-tensor-cd}\cite{neud,commut,tracy-jina}
	Let $c$ and $c'$ of size $n^2\times p^2$ and $m^2\times q^2$ respectively,  with a canonical block partition (into  blocks of the same size $n\times p$ and $m\times q$ respectively). 	Let $K_{mn}$ denote the  commutation matrix of size $mn$. Then		\begin{equation}\label{eqn-formula-ts-tensor-general}
	c \boxtimes c'\,=\, 	(\mat{I}_{n}\,\otimes K_{mn} \,\otimes \mat{I}_{m})\,	\cdot\, (c\otimes c')\, \cdot\, (\mat{I}_{p}\,\otimes K_{pq}\,\otimes  \mat{I}_{q})
	\end{equation}
	\end{prop}
If  $c$ and $c'$  are of size $n^2\times p^2$ and $m^2\times q^2$ respectively,  with the canonical block partition,   then  their  Tracy-Singh product represents the linear transformation  $F_{23}\,(c \otimes c')\,F_{23}\,:\,\mathbb{C}^n \otimes \mathbb{C}^m \otimes \mathbb{C}^n \otimes \mathbb{C}^m\,\rightarrow\mathbb{C}^p \otimes \mathbb{C}^q \otimes \mathbb{C}^p\otimes \mathbb{C}^q$, where  $F_{23}$  exchanges the two middle factors \cite{chou-quantum}. In \cite{chou-quantum}, it is shown that   in the case that $c$ and $c'$ are solutions of the YBE,  the  Tracy-Singh product coincides with  the operation called   \emph{the tensor product of $R$-matrices} (although it differs from the actual tensor product $\otimes$) in  \cite{gandal1,gandal2}. In \cite{majid-book}, S. Majid refers to the cabling operation  of $R$-matrices.
\section{Realization of  the Yang-Baxter  gates}
\subsection{Realization of  two-qudit Yang-Baxter  gates}
A \emph{$n$-qubit} is a state vector in the Hilbert space $(\mathbb{C}^2)^{\otimes n}$ and a \emph{$n$-qudit} is a state vector in the Hilbert space $(\mathbb{C}^d)^{\otimes n}$.  A $n$-qudit $\mid \phi \rangle$ is   \emph{decomposable} if $\mid \phi \rangle\,=\,\mid \phi_1 \rangle\,\otimes\,\mid \phi_2 \rangle\,\otimes\,...\otimes\,\mid \phi_n \rangle$, where 
$\mid \phi_i \rangle\,\in \, \mathbb{C}^d$, for $ 1 \leq i \leq n$.  Otherwise, $\mid \phi \rangle$ is  \emph{entangled}.  A \emph{(quantum)  $n$-qudits gate} is a unitary operator  $L:\,(\mathbb{C}^d)^{\otimes n}\,\rightarrow\,(\mathbb{C}^d)^{\otimes n}$. These gates belong to the unitary group $\operatorname{U}((\mathbb{C}^d)^{\otimes n})$. A sequence $L_1,L_2,...,L_k$ of quantum gates constitutes a \emph{quantum circuit} on $n$-qudits, which output is the product gate  
$L_1\cdot L_2\cdot...\cdot L_k$. In practice, one wants to build circuits out of gates which are local in the sense that they operate on a small number of qudits: $1$, $2$, or $3$ \cite{bryl}. 
\begin{defn}\cite{bryl} A collection $\mathcal{U}$  of $1$-qudit gates $\{S_i\}$ and $2$-qudit gates $\{U_j\}$ is called \emph{universal} if, for each $n \geq  2$, every $n$-qudit gate can be approximated with arbitrary accuracy by a circuit made up of the $n$-qudit gates produced by the $\{S_i\}$ and $\{U_j\}$, and it  is called \emph{exactly universal} if, for each $n \geq  2$, every $n$-qudit gate can be obtained exactly by a circuit made up of the $n$-qudit gates produced by the $\{S_i\}$ and $\{U_j\}$.
\end{defn}A $2$-qudit gate $L:\,(\mathbb{C}^d)^{\otimes 2}\,\rightarrow\,(\mathbb{C}^d)^{\otimes 2}$ is \emph{primitive} if $L$ maps decomposable $2$-qudit to decomposable $2$-qudit, otherwise $L$ is  said to be \emph{imprimitive} in  \cite{bryl} or \emph{entangling} in \cite{kauf-lo3}.  In \cite{bryl}, it  is  proved that $L$ is entangling if and only if  $L$ is exactly universal, which means  that the collection of all $1$-qudit gates together with $L$ generates the unitary group  $\operatorname{U}((\mathbb{C}^d)^{\otimes n})$, for every $n \geq 2$.  There is a criteria to determine whether a $2$-qudit is primitive \cite{bryl}.
\begin{ex}\label{ex-cnot-2qubit}
	A very important example of  entangling 2-qubit   gate is the following   unitary matrix of size $4$ which acts on $2$-qubits: 
	$CNOT=\begin{pmatrix}
	1& 0 & 0 & 	0\\
	0 & 1& 0 & 0 \\
	0 &	0 & 0& 1 \\
	0&0 & 1 &	0\\
	\end{pmatrix}$;  not an $R$-matrix.
\end{ex} 
\subsection{Realization of  four-qudit  gates coming from the Tracy-Singh product}
  The following fact is a direct application of  Proposition \ref{prop-box-cd-tensor-cd}.
 \begin{prop}\label{theo-tracy=swap-tensor}
 	Let $c,c',\,P:\,(\mathbb{C}^d)^{\otimes 2}\,\rightarrow\,(\mathbb{C}^d)^{\otimes 2}$  be $2$-qudit gates, such that $P$ is the swap map. Let $c\boxtimes c':\,(\mathbb{C}^d)^{\otimes 4}\,\rightarrow\,(\mathbb{C}^d)^{\otimes 4}$  be the $4$-qudit gate defined by their Tracy-Singh product $c \boxtimes c'$   with the canonical block partition. Then $c\boxtimes c'\,=\,( \mat{I}_{d}\otimes P\otimes \mat{I}_{d})\,(c\otimes c')\,( \mat{I}_{d}\otimes P\otimes \mat{I}_{d})$:
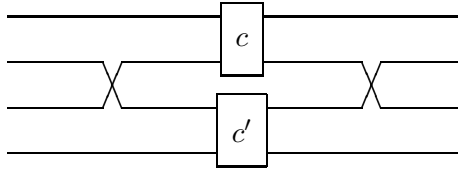
 	\[	\Qcircuit @C=0.25cm @R=.25cm {
 		& \qw &\qw & \qw & \qw & \qw & \qw&\qw & \qw & \qw & \qw &\multigate{1}{c}& \qw& \qw & \qw & \qw & \qw & \qw & \qw & \qw & \qw& \qw\\
 		& \qw & \qw & \qw& \qw &\qw\link{1}{1} & \link{1}{-1} & \qw & \qw & \qw & \qw & \ghost{c} &\qw &\qw&\qw& \qw& \qw\link{1}{1} & \link{1}{-1} & \qw & \qw & \qw& \qw  \\
 		& \qw & \qw & \qw& \qw &\qw\link{-1}{1} & \link{-1}{-1} & \qw& \qw & \qw & \qw &\multigate{1}{c'}  & \qw & \qw              & \qw& \qw &\qw\link{-1}{1} & \link{-1}{-1} & \qw & \qw & \qw & \qw  \\
 		& \qw & \qw & \qw & \qw & \qw & \qw&\qw& \qw & \qw & \qw &\ghost{c'} & \qw& \qw    & \qw & \qw & \qw& \qw   & \qw &  \qw& \qw & \qw\\
 	}\]
\end{prop}
Since for every integer $d\geq 2$,  there exists an entangling  $2$-qudit gate  $R:\,(\mathbb{C}^d)^{\otimes 2}\,\rightarrow\,(\mathbb{C}^d)^{\otimes 2}$, where  $R$  satisfies  the YBE \cite{chou-quantum}, 
 one can assume  that for every $d\geq 2$,  there exists an exactly universal set of gates $\mathcal{U}_d$  with a single $2$-qudit gate $U$. We prove now Theorem \ref{theorem-paper}, that is we show that, whenever the realisation of the gates $c,c':\,(\mathbb{C}^d)^{\otimes 2}\,\rightarrow\,(\mathbb{C}^d)^{\otimes 2}$ in terms of gates from $\mathcal{U}_d$ is given, there exists a simple way to realise  any  $4$-qudit gate of the form  $c \boxtimes c'$  in terms of  the gates from  $\mathcal{U}_d$. And in the particular case that  $c \boxtimes c'$ acts on a Hilbert space of dimension $2^{\ell}$,  $U$ can be taken to be the CNOT  gate. 
 \begin{proof}[Proof of Theorem \ref{theorem-paper}]
 	Let 	$c, c',\,L_1,...,L_k,\,M_1,...,M_l :\, (\mathbb{C}^d)^{\otimes 2} \rightarrow (\mathbb{C}^d)^{\otimes 2}$  be $2$-qudit gates, such that $c\,=\,L_1L_2...L_k$ and $c'\,=\,M_1M_2...M_l$. 
 Let $\mathcal{U}_d$ be a universal set of gates,  with  $U$ the single  $2$-qudit gate.  Let $S_i,T_i,S'_j,T'_j\,:\,\mathbb{C}^d\,\rightarrow\,\mathbb{C}^d$, $1\leq i \leq k$, $1 \leq j \leq l$,  be $1$-qudit gates from $\mathcal{U}_d$,  or maybe $I_{d}$, such   that $L_i\,=\,S_i\otimes T_i$ or $L_i=U$ and  $M_j\,=\,S'_j\otimes T'_j$ or $M_j=U$.  Assuming with n.l.o.g that $k \leq l$, from the properties of $\otimes$ and Theorem \ref{thm-tracy}$(v)$:
		\begin{equation}\label{eqn-product of tensor}
		c \otimes c'\,=\, L_1L_2...L_k\,\otimes\,M_1M_2...M_l\,=\, (L_1\,\otimes\,M_1)\,(L_2\,\otimes\,M_2)...(L_k\,\otimes\,M_k)...(I_{d^2}\,\otimes\,M_l)	\end{equation}
			\begin{equation}\label{eqn-tracy-1-gates}		
		c \boxtimes c'\,=\, L_1L_2...L_k\,\boxtimes\,M_1M_2...M_l\,=\, (L_1\,\boxtimes\,M_1)\,(L_2\,\boxtimes\,M_2)...(L_k\,\boxtimes\,M_k)...(I_{d^2}\,\boxtimes\,M_l)	\end{equation}
From Equations (\ref{eqn-product of tensor})-(\ref{eqn-tracy-1-gates}),  knowing the realisation  of the gates $(L_i\,\otimes\,M_i)$ or $(L_i\,\boxtimes\,M_i)$
in terms of   gates from $\mathcal{U}_d$ gives  a  realisation of  $c\boxtimes c'$ of  the following form:
\begin{gather}\label{eqn-tracy-1-gates2}		
c \boxtimes c'\,=\, ( \mat{I}_{d}\otimes P\otimes \mat{I}_{d})\,\,(\prod\limits_{i=1}^{l}L_i\otimes M_i)\,\,\,( \mat{I}_{d}\otimes P\otimes \mat{I}_{d})
\end{gather}
where for every $1 \leq i \leq l$, $L_i\otimes M_i$ has one of the following forms, where $S_i, T_i,S'_i, T'_i$ may be equal to $I_d$: $S_i\otimes T_i\otimes S'_i\otimes T'_i \;\;\textrm{or} \;\;
U\otimes S'_i\otimes T'_i  \;\;\textrm{or}\;\;
S_i\otimes T_i\otimes U \;\;\textrm{or}\;\;
U\otimes U$. For many special cases a simpler form for $c \boxtimes c'$ can be achieved. 
\end{proof}
\section{Proof of Theorem \ref{theorem-paper2}}
Let $c$   be a  YBE  gate and let  the family of representations of $B_n$ obtained from $c$:

\begin{gather}
\rho_n^{c}:\,B_n\rightarrow \operatorname{GL}((\mathbb{C}^d)^{\otimes n})\\
\sigma_j \mapsto (Id_d)^{\otimes\, j-1}\,\, \otimes \,\, c \,\, \otimes \,\,(Id_d)^{\otimes\, n- j-1} \nonumber
\end{gather}
where $\sigma_{1},\sigma_{2},...,\sigma_{n-1}$ are generators of $B_n$. Let $\mu \in  \operatorname{End}(V)$. A pair $(c,\mu)$ is an \emph{enhanced YBE pair}
if $c$ commutes with $\mu \otimes\mu$ and $\operatorname{Tr}_2(c\circ\mu^{\otimes 2} )\,=\,\operatorname{Tr}_2(c^{-1}\circ \mu^{\otimes 2})\,=\,\mu$, where $\operatorname{Tr}_2$ denotes the partial trace over the second factor.  Since  any oriented link is equivalent (ambient isotopic) to the trace closure of some braid $b$,    for $(c,\mu)$  an enhanced YBE pair \begin{gather}
I_{c}(b)\,=\,\operatorname{Tr}(\rho_n^{c}(b)\circ\mu^{\otimes n})\end{gather}
 can be defined for every $b \in B_n$  and it yields a link invariant \cite{turaev}.
\begin{lem}\label{lemma-enhancedybe}
Let 	 $(c,\mu)$ and $(c',\eta)$ be two enhanced  YBE pairs. Let $c \boxtimes c'$ their  Tracy-Singh product with the  canonical block partition. Then $(c \boxtimes c', \mu \otimes \eta)$ is  an enhanced  YBE pair.
\end{lem}
\begin{proof}
	First, we show that  $c \boxtimes c'$ commutes with $(\mu \otimes \eta)^{\otimes 2}$.  
In the appendix, we  prove  that with the canonical partition of $\mu \otimes \mu$ and $\eta\otimes \eta$ into blocks, Equation (\ref{eqn-mu-eta2}) holds:
\begin{equation}\label{eqn-mu-eta2}
(\mu \otimes \eta)^{\otimes 2}\,=\,(\mu \otimes \mu )\,\boxtimes (\eta\otimes \eta)
\end{equation}
So, $(c \boxtimes c')\,\cdot\,(\mu \otimes \eta)^{\otimes 2}\,=\,(c \boxtimes c')\,\cdot\;((\mu \otimes \mu )\,\boxtimes (\eta\otimes \eta))\,=\,(c(\mu \otimes \mu ))\boxtimes(c'(\eta\otimes \eta))$, from Theorem \ref{thm-tracy}$(v)$. As $c$ commutes with $\mu \otimes \mu $ and $c'$ commutes with $\eta\otimes \eta$, this is equal to  $(\mu \otimes \eta)^{\otimes 2} \,\c(c \boxtimes c')$. Next, we show  $\operatorname{Tr}_2((c\boxtimes c')\circ (\mu\otimes \eta)^{\otimes 2})\,=\,\operatorname{Tr}_2((c\boxtimes c')^{-1}\circ (\mu\otimes \eta)^{\otimes 2})\,=\,\mu \otimes \eta$. As above, considering the product of the corresponding matrices, 
$(c\boxtimes c')\cdot (\mu\otimes \eta)^{\otimes 2})\,=\, (c(\mu \otimes \mu ))\boxtimes(c'(\eta\otimes \eta))$ and $(c\boxtimes c')^{-1}\cdot (\mu\otimes \eta)^{\otimes 2}\,=\,(c^{-1}(\mu \otimes \mu ))\boxtimes(c'^{-1}(\eta\otimes \eta))$, using Theorem \ref{thm-tracy}$(vii)$.
We show in the appendix that if $A$ and $B$ are both square matrices of order $d^2$, with the canonical partition into blocks, then:  
\begin{equation}\label{eqn-general}
\operatorname{Tr}_2(A\boxtimes B)\,=\,\operatorname{Tr}_2(A)\,\otimes \, \operatorname{Tr}_2(B)
\end{equation}
So,  since $(c,\mu)$ and $(c',\eta)$ are enhanced  YBE pairs, we have from Equation (\ref{eqn-general}):
\begin{gather*}
\operatorname{Tr}_2((c^{\pm 1}\boxtimes c'^{\pm 1})\cdot (\mu\otimes \eta)^{\otimes 2})\,=\,\operatorname{Tr}_2((c^{\pm 1}(\mu \otimes \mu ))\boxtimes(c'^{\pm 1}(\eta\otimes \eta)))\\= \operatorname{Tr}_2(c^{\pm 1}(\mu \otimes \mu ))\,\otimes\,\operatorname{Tr}_2(c'^{\pm 1}(\eta\otimes \eta))\,=\,\mu\,\otimes \eta
\end{gather*}
That is, the pair $(c \boxtimes c', \mu \otimes \eta)$ is  an enhanced  YBE pair.
 \end{proof}
\begin{proof}[Proof of Theorem \ref{theorem-paper2}]
	From Lemma \ref{lemma-enhancedybe}, the pair $(c \boxtimes c', \mu \otimes \eta)$ is  an enhanced  YBE pair. Assume  that $c$ and $c'$ are primitive,   $c\,=\,c_1\otimes c_2$ and  $c'\,=\,d_1\otimes d_2$. 	We  prove that $I_{c\boxtimes c'}$ is equal to the product of $I_{c}$ and $I_{c'}$. We define:
\begin{gather*}
\rho_n^{c\boxtimes c'}:\,B_n\rightarrow \operatorname{GL}((\mathbb{C}^{d^2})^{\otimes n})\\
\sigma_j \mapsto (Id_{d^2})^{\otimes\, j-1}\,\, \otimes \,\, (c\boxtimes c')\,\,\otimes\,\, (Id_{d^2})^{\otimes\, n- j-1}
\end{gather*}
From \cite[Theorem 6.3]{chou-quantum},   
$c \boxtimes d\,=\,(c_1\otimes d_1)\,\otimes \,(c_2\otimes d_2)$, so:
\begin{gather*}
\rho_n^{c}(\sigma_j)\,=\,(Id_{d})^{\otimes\, j-1}\,\, \otimes \,\, c_1 \otimes c_2\,\,\otimes\,\, (Id_{d})^{\otimes\, n- j-1} 
\\
\rho_n^{c\boxtimes c'}(\sigma_j)=\,(Id_{d^2})^{\otimes\, j-1}\,\, \otimes \,\, c_1\otimes d_1\,\otimes \,c_2\otimes d_2\,\,\otimes\,\, (Id_{d^2})^{\otimes\, n- j-1} \nonumber
\end{gather*}
As $\rho_n^{c}$ and $\rho_n^{c\boxtimes c'}$ are homomorphisms of groups, for $b= \sigma_{i_1}^{m_1}\sigma_{i_2}^{m_2}...\sigma_{i_k}^{m_k}\,\in B_n$, $\rho_n^{c}(b)$ and $\rho_n^{c\boxtimes c'}(b)$ have the following schematic form, where the exponents $m_{.}$ may be equal zero:
\begin{gather*}
\rho_n^{c}(b)\,=\,(Id_{d})^{\otimes\, \cdot}\,\, \otimes \,\, \prod\limits_{j=1}^{}c_1^{m_{j_1}}  c_2^{m_{j_2}}\,\,\otimes ...\otimes \,\, \prod\limits_{l=1}^{}c_1^{m_{l_1}} \, c_2^{m_{l_2}}\,\,\otimes\,\, (Id_{d})^{\otimes\, \cdot\cdot} \\
\rho_n^{c\boxtimes c'}(b)=\,(Id_{d^2})^{\otimes\, \cdot}\,\,\, \otimes \, \prod\limits_{j=1}^{}c_1^{m_{j_1}}  c_2^{m_{j_2}}\, \,\otimes\,\, \prod\limits_{j=1}^{}d_1^{m_{j_3}} \, d_2^{m_{j_4}}\,\otimes ..\otimes \, \prod\limits_{l=1}^{}c_1^{m_{l_1}} \, 
c_2^{m_{l_2}}\, \, \,\otimes\,\, \prod\limits_{j=1}^{}d_1^{m_{l_3}} \, d_2^{m_{l_4}}\,\otimes\,(Id_{d^2})^{\otimes\, \cdot\cdot} 
\end{gather*}
We compute  $I_{c}(b)$ and $I_{c'}(b)$,  $b \in B_n$:
\begin{gather*}
I_{c}(b)\,=\, \operatorname{Tr}(\,\rho_n^{c}(b)\cdot(\mu)^{\otimes n}\,)\,
=\,\operatorname{Tr}(\,\,((Id_{d})^{\otimes\, \cdot}\,\, \otimes \,\, \prod\limits_{j=1}^{}c_1^{m_{j_1}}  c_2^{m_{j_2}}\,\,\otimes ...\otimes \,\, \prod\limits_{l=1}^{}c_1^{m_{l_1}} \, c_2^{m_{l_2}}\,\,\otimes\,\, (Id_{d})^{\otimes\, \cdot\cdot}) \,\cdot(\mu)^{\otimes n}\,)\\
=\operatorname{Tr}(\,\,(\mu)^{\otimes\, \cdot}\,\, \otimes \,\, (\prod\limits_{j=1}^{}c_1^{m_{j_1}}  c_2^{m_{j_2}})\,\mu\,\otimes ...\otimes \,\, (\prod\limits_{l=1}^{}c_1^{m_{l_1}} \, c_2^{m_{l_2}})\,\mu\,\otimes\,\, (\mu)^{\otimes\, \cdot\cdot} \,)\\
=\operatorname{Tr}(\mu)^{\cdot}\,\,\operatorname{Tr}(\,(\prod\limits_{j=1}^{}c_1^{m_{j_1}}  c_2^{m_{j_2}})\,\mu\,)...
\operatorname{Tr}(\,(\prod\limits_{l=1}^{}c_1^{m_{l_1}} \, c_2^{m_{l_2}})\,\mu\,)\, \operatorname{Tr}(\mu)^{\cdot\cdot}\\
\textrm{\,In\ the\,same\,way, \,\,}
I_{c'}(b)\,=\,\operatorname{Tr}(\eta)^{\cdot}\,\,\operatorname{Tr}(\,(\prod\limits_{j=1}^{}d_1^{m_{j_3}}  d_2^{m_{j_4}})\,\eta\,)...
\operatorname{Tr}(\,(\prod\limits_{l=1}^{}d_1^{m_{l_3}} \, d_2^{m_{l_4}})\,\eta\,)\, \operatorname{Tr}(\eta)^{\cdot\cdot}
\end{gather*}
 We compute  $I_{c\boxtimes c'}(b)$, $b \in B_n$:
 \begin{gather*}
 I_{c\boxtimes c'}(b)\,=\, \operatorname{Tr}(\,\rho_n^{c\boxtimes c'}(b)\cdot(\mu \otimes\eta)^{\otimes n}\,)\\
 =\operatorname{Tr}(\,(\mu \otimes\eta)^{\otimes \cdot}\,\,\otimes \, \prod\limits_{j=1}^{}(c_1^{m_{j_1}}  c_2^{m_{j_2}})\,\mu\,\,\otimes (\prod\limits_{j=1}^{}d_1^{m_{j_3}} \, d_2^{m_{j_4}}\,)\,\eta\,\otimes ...\otimes\,(\prod\limits_{j=1}^{}d_1^{m_{l_3}} \, d_2^{m_{l_4}})\eta\,\otimes (\mu \otimes\eta)^{\otimes \cdot \cdot})\\
 =\operatorname{Tr}(\,\mu \,)^{\cdot}\,\operatorname{Tr}(\,\eta)\,)^{\cdot}\,\,\operatorname{Tr}(\, (\prod\limits_{j=1}^{}c_1^{m_{j_1}} c_2^{m_{j_2}})\, \mu\,)\,.. \, \operatorname{Tr}(\,(\prod\limits_{l=1}^{}d_1^{m_{l_3}} 
 \, d_2^{m_{l_4}})\,\eta\,)\,\operatorname{Tr}(\,\mu \,)^{\cdot \cdot}\,\operatorname{Tr}(\,\eta\,)^{\cdot \cdot}
 \end{gather*}
 So,  $I_{c\boxtimes c'}(b)\,=\, I_{c}(b)\cdot I_{c'}(b)$.
\end{proof}
If  $c$ and $c'$ are primitive of the form  $c\,=\,(c_1\otimes c_2)P$ and  $c'\,=\,(d_1\otimes d_2)P$, where $P$ is  the  swap map on $(\mathbb{C}^d)^{\otimes 2}$, then from \cite[Theorem 6.3]{chou-quantum},  
$c \boxtimes d\,=\,((c_1\otimes d_1)\,\otimes \,(c_2\otimes d_2))\,Q$, where $Q$ is  the  swap map on $(\mathbb{C}^{d^2})^{\otimes 2}$. In that case, we could not prove, using the same kind of computation as in the proof of Theorem \ref{theorem-paper2}, that $I_{c\boxtimes c'}(b)\,=\, I_{c}(b)\cdot I_{c'}(b)$, although it seems that is should be true.

\section{Appendix: Proof of Equations   (\ref{eqn-mu-eta2}) and (\ref{eqn-general})}
\subsection*{ Proof of Equation  (\ref{eqn-mu-eta2})}
We  prove  that with the canonical partition of $\mu \otimes \mu$ and $\eta\otimes \eta$ into blocks, 
$(\mu \otimes \eta)^{\otimes 2}\,=\,(\mu \otimes \mu )\,\boxtimes (\eta\otimes \eta)$:
\begin{gather*}
(\mu \otimes \mu )\,\boxtimes (\eta\otimes \eta)=
\begin{pmatrix}
\mu_{11}\,\mu
& \rvline & \mu_{12}\,\mu
& \rvline &... &\rvline & \mu_{1d}\,\mu\\
\hline
...
& \rvline & ...
& \rvline &... &\rvline & ... \\
\hline
\mu_{d1}\,\mu
& \rvline & \mu_{d2}\,\mu
& \rvline &... &\rvline & \mu_{dd}\,\mu \\
\end{pmatrix}
\boxtimes
\begin{pmatrix}
\eta_{11}\,\eta
& \rvline & \eta_{12}\,\eta
& \rvline &... &\rvline & \eta_{1d}\,\eta\\
\hline
...
& \rvline & ...
& \rvline &... &\rvline & ... \\
\hline
\eta_{d1}\,\eta
& \rvline & \eta_{d2}\,\eta
& \rvline &... &\rvline & \eta_{dd}\,\eta\\
\end{pmatrix}\\
\vspace*{0.9cm}
=\,\begin{pmatrix}
\mu_{11}  \eta_{11}(\mu \otimes \eta)
..
&.. \mu_{11}  \eta_{1d}(\mu \otimes \eta)&\rvline &.. &\rvline& 	\mu_{1d}  \eta_{11}(\mu \otimes \eta).. & ..	\mu_{1d}  \eta_{1d}(\mu \otimes \eta)\\
\hdashline[2pt/2pt]
...
& ..
.&\rvline &.. &\rvline& ...&... \\
\hdashline[2pt/2pt]
\mu_{11}  \eta_{d1}(\mu \otimes \eta)
..
&.. \mu_{11}  \eta_{dd}(\mu \otimes \eta)&\rvline &.. &\rvline& 	\mu_{1d}  \eta_{d1}(\mu \otimes \eta).. & ..	\mu_{1d}  \eta_{dd}(\mu \otimes \eta)\\
\hline
...
& .
..&\rvline &.. &\rvline& ...&... \\
\hdashline[2pt/2pt]
\mu_{d1}  \eta_{d1}(\mu \otimes \eta)
..
&.. \mu_{d1}  \eta_{dd}(\mu \otimes \eta)&\rvline &.. &\rvline& 	\mu_{dd}  \eta_{d1}(\mu \otimes \eta).. & ..	\mu_{dd}  \eta_{dd}(\mu \otimes \eta)\\
\end{pmatrix}\\
\vspace*{0.3cm}
=\,(\mu \otimes \eta)\,\,\otimes\,\,(\mu \otimes \eta)
\end{gather*}
\subsection*{ Proof of Equation  (\ref{eqn-general})}
Let $A$ and $B$ be  square matrices of order $d^2$, with the canonical partition into blocks. We show that  $\operatorname{Tr}_2(A\boxtimes B)\,=\,\operatorname{Tr}_2(A)\,\otimes \, \operatorname{Tr}_2(B)$. 
\begin{gather*}
\operatorname{Tr}_2(A \boxtimes B)=
\operatorname{Tr}_2\big(\begin{pmatrix}
A_{11}
& \rvline & A_{12}
& \rvline &... &\rvline & A_{1d} \\
\hline
...
& \rvline & ...
& \rvline &... &\rvline & ... \\
\hline
A_{d1}
& \rvline & A_{d2}
& \rvline &... &\rvline & A_{dd} \\
\end{pmatrix}
\boxtimes
\begin{pmatrix}
B_{11}
& \rvline & B_{12}
& \rvline &... &\rvline & B_{1d} \\
\hline
...
& \rvline & ...
& \rvline &... &\rvline & ... \\
\hline
B_{d1}
& \rvline & B_{d2}
& \rvline &... &\rvline & B_{dd} \\
\end{pmatrix}\big)\\
	\end{gather*}
\begin{gather*}
=\,\operatorname{Tr}_2\,\begin{pmatrix}
A_{11}\otimes  B_{11}
..
&.. A_{11}\otimes B_{1d}&\rvline &.. &\rvline& 	A_{1d}\otimes B_{11}.. & ..	A_{1d}\otimes B_{1d} \\
\hdashline[2pt/2pt]
...
& ..
.&\rvline &.. &\rvline& ...&... \\
\hdashline[2pt/2pt]
A_{11}\otimes B_{d1}
..
&..A_{11}\otimes B_{dd} &\rvline &.. &\rvline& A_{1d}\otimes B_{d1} ..&	..A_{1d}\otimes B_{dd} \\
\hline
...
& .
..&\rvline &.. &\rvline& ...&... \\
\hdashline[2pt/2pt]
A_{d1}\otimes B_{d1}
..
&..A_{d1}\otimes B_{dd}&\rvline &.. &\rvline& A_{dd}\otimes B_{d1}..&..	A_{dd}\otimes B_{dd} \\
\end{pmatrix}\\
	\vspace*{0.9cm}
=\,\begin{pmatrix}
\operatorname{Tr}(A_{11}\otimes  B_{11})
..
&.. \operatorname{Tr}(A_{11}\otimes B_{1d})&\rvline &.. &\rvline& 	\operatorname{Tr}(A_{1d}\otimes B_{11}).. & ..	\operatorname{Tr}(A_{1d}\otimes B_{1d}) \\
\hdashline[2pt/2pt]
...
& ..
.&\rvline &.. &\rvline& ...&... \\
\hdashline[2pt/2pt]
\operatorname{Tr}(A_{11}\otimes B_{d1})
..
&..\operatorname{Tr}(A_{11}\otimes B_{dd})&\rvline &.. &\rvline& \operatorname{Tr}(A_{1d}\otimes B_{d1}) ..&	..\operatorname{Tr}(A_{1d}\otimes B_{dd})\\
\hline
...
& .
..&\rvline &.. &\rvline& ...&... \\
\hdashline[2pt/2pt]
\operatorname{Tr}(A_{d1}\otimes B_{d1})
..
&..\operatorname{Tr}(A_{d1}\otimes B_{dd})&\rvline &.. &\rvline& \operatorname{Tr}(A_{dd}\otimes B_{d1})..&..	\operatorname{Tr}(A_{dd}\otimes B_{dd}) \\
\end{pmatrix}\\
	\vspace*{0.5cm}
=\,\begin{pmatrix}
\operatorname{Tr}(A_{11}) \operatorname{Tr }(B_{11})
..
&.. \operatorname{Tr}(A_{11})\operatorname{Tr}( B_{1d})&\rvline &.. &\rvline& 	\operatorname{Tr}(A_{1d})\operatorname{Tr}( B_{11}).. & ..	\operatorname{Tr}(A_{1d})\operatorname{Tr}(B_{1d}) \\
\hdashline[2pt/2pt]
...
& ..
.&\rvline &.. &\rvline& ...&... \\
\hdashline[2pt/2pt]
\operatorname{Tr}(A_{11})\operatorname{Tr}(B_{d1})
..
&..\operatorname{Tr}(A_{11})\operatorname{Tr}( B_{dd})&\rvline &.. &\rvline& \operatorname{Tr}(A_{1d})\operatorname{Tr}(B_{d1}) ..&	..\operatorname{Tr}(A_{1d})\operatorname{Tr}( B_{dd})\\
\hline
...
& .
..&\rvline &.. &\rvline& ...&... \\
\hdashline[2pt/2pt]
\operatorname{Tr}(A_{d1})\operatorname{Tr}( B_{d1})
..
&..\operatorname{Tr}(A_{d1})\operatorname{Tr}(B_{dd})&\rvline &.. &\rvline& \operatorname{Tr}(A_{dd})\operatorname{Tr}( B_{d1})..&..	\operatorname{Tr}(A_{dd})\operatorname{Tr}( B_{dd}) \\
\end{pmatrix}\\
\vspace*{0.9cm}
=\,\begin{pmatrix}
\operatorname{Tr }(A_{11})&
..
&.. &\operatorname{Tr}( A_{1d})\\
\hdashline[2pt/2pt]
...
& ..&..&..\\
\hdashline[2pt/2pt]
\operatorname{Tr }(A_{d1})&
..
&.. &\operatorname{Tr}( A_{dd})\\
\end{pmatrix}\,
\otimes\,
\begin{pmatrix}
 \operatorname{Tr }(B_{11})&
..
&.. &\operatorname{Tr}( B_{1d})\\
\hdashline[2pt/2pt]
...
& ..&..&..\\
\hdashline[2pt/2pt]
 \operatorname{Tr }(B_{d1})&
..
&.. &\operatorname{Tr}( B_{dd})\\
\end{pmatrix}\,
=\,\operatorname{Tr}_2( A)\,\otimes\,\operatorname{Tr}_2( B)
\end{gather*}


\bigskip\bigskip\noindent
\smallskip\noindent

\smallskip\noindent

\end{document}